\documentclass[twoside]{article}
\usepackage{fleqn,espcrc2}

\usepackage{epsfig}
\usepackage[figuresright]{rotating}

\newcommand{\AmS}{{\protect\the\textfont2
  A\kern-.1667em\lower.5ex\hbox{M}\kern-.125emS}}

\title{Determinations of $|V_{ub}|$ with inclusive techniques at LEP}

\author{M. Battaglia\address{Dept. of Physics,
        University of Helsinki, \\ 
        FIN-00014 Helsinki, Finland}%
        \thanks{present address: CERN/EP Division, CH-1211 Geneva}}

\begin{document}

\begin{abstract}

The charmless semileptonic decay $B$ branching fraction has been measured, 
using inclusive techniques, by the {\sc Aleph}, {\sc Delphi} and L3 
experiments at LEP. The average of their results is
\begin{center}
BR($b \rightarrow X_u \ell \bar \nu$) = 
(1.74 $\pm$ 0.37 (stat.+exp.) $\pm$ 0.38 ($b \rightarrow c$) 
$\pm$ 0.21 ($b \rightarrow u$)) $\times$ 10$^{-3}$. 
\end{center}
From this result the value of the $|V_{ub}|$ element in the CKM mixing matrix 
has been derived, using OPE predictions, obtaining:
\begin{center}
$|V_{ub}| = (4.13 ^{+0.42}_{-0.47}~\mathrm{(stat.+det.)} 
      ^{+0.43}_{-0.48}~\mathrm{(b \rightarrow c~syst.)} 
      ^{+0.24}_{-0.25}~\mathrm{(b \rightarrow u~sys.)}$
      $\pm 0.02~(\tau_b) 
      \pm 0.20~\mathrm{(HQE)}) \times 10^{-3}$.
\end{center}

\vspace{1pc}
\end{abstract}

\maketitle

\section{INTRODUCTION}

The accuracy in the determination of the $|V_{ub}|$ element in the 
Cabibbo-Kobayashi-Maskawa mixing matrix~\cite{ckm} plays an important role 
in the study of the unitarity triangle and in the related tests of the 
Standard Model (SM). In particular, a non-vanishing $|V_{ub}|$ is 
essential to preserve the possibility to describe CP violation within the SM 
and its value places a direct constraint on the magnitude of the CP violating 
phase $\beta$.
 
The first determination of the magnitude of $|V_{ub}|$ was obtained from the 
yield of leptons produced with momentum above the kinematic 
limit for $b \rightarrow X_c \ell \bar \nu$ transitions, first reported by 
{\sc Cleo}~\cite{cleo1} and soon confirmed by {\sc Argus}~\cite{argus1}. 
However this method is sensitive to only $\simeq$~10\% of the inclusive
charmless semileptonic (s.l.) yield, and the extraction of $|V_{ub}|$ is 
subject to a large model dependence. 

More recently, exclusive 
$B \rightarrow \pi \ell \bar \nu$ and $B \rightarrow \rho \ell \bar \nu$ 
decays have been measured by {\sc Cleo}~\cite{cleo2}. The determination of 
$|V_{ub}|$ from exclusive s.l. decays has still a significant model 
dependence. First lattice estimates of the relevant form factors indicate that
a significant reduction of these uncertainties may be expected in the future.

The extraction of $|V_{ub}|$ from the distribution of the invariant mass $M_X$
of the hadronic system recoiling against the lepton pair peaked, for
$b \rightarrow X_u \ell \bar \nu$, at a significantly lower value  than for
$b \rightarrow X_c \ell \bar \nu$ was proposed several years 
ago~\cite{barger}, and it has recently been the subject of new theoretical 
calculations~\cite{falk,bdu}. If $b \rightarrow u$ transitions can be 
discriminated from the dominant $b \rightarrow c$ background up to 
$M_X \simeq M(D)$, this method is sensitive to $\simeq$~80\% of the charmless
s.l. $B$ decay rate. Further, if no preferential weight is given to 
low mass states in the event selection, the non-perturbative effects 
are expected to be small and the OPE description of the transition has been 
shown to be accurate away from the resonance region. 

The experimental 
challenge comes from the requirement to isolate the $b \rightarrow u$ 
contribution to the s.l. yield from the $\simeq$~60 times larger 
$b \rightarrow c$ one while ensuring a uniform sampling of the decay phase 
space to avoid biases towards a few exclusive low-mass, low-multiplicity 
states, such as $\pi \ell \bar \nu$ and $\rho \ell \bar \nu$. 
In principle, this 
method is well suited for the LEP 
experiments, where the recorded statistics is not sufficient for studying the 
exclusive decay modes with good accuracy while the
significant boost of the $B$ hadrons, the separation of $b$ and $\bar b$ decay
products in opposite hemispheres, and the good secondary vertex reconstruction 
capabilities make the study of inclusive decay $B$ decays possible. 
The analysis techniques adopted by the {\sc Aleph}~\cite{aleph_vub}, 
{\sc Delphi}~\cite{delphi_vub} and L3~\cite{l3_vub} Collaborations are based 
on the observation that 
$b \rightarrow X_u \ell \bar \nu$ decays can be inclusively discriminated from
$b \rightarrow X_c \ell \bar \nu$ by exploiting
the differences in the invariant mass and kaon content of the secondary 
hadronic system, in the decay multiplicity and in the decay vertex topology.
These features have been used differently in the three analyses, resulting in 
determinations of the charmless s.l. branching fraction obtained from 
samples with varying efficiency and purity and with systematic uncertainties 
that are only partially correlated. 

\section{EXPERIMENTAL RESULTS}

The analyses consist of three main steps: i) inclusive selection 
of s.l. $B$ decay candidates, ii) definition of a subsample enriched in
$b \rightarrow u$ transitions and iii) measurement of the charmless
s.l. branching fraction. All experiments started from a sample of events 
containing an identified electron or muon selected from a combined data set 
of $\simeq$ 8.2~M hadronic $Z^0$ decays. {\sc Aleph} and {\sc Delphi} also 
imposed $b$-tag criteria to reject light quark and charm backgrounds.  
\begin{figure}[h!]

\vspace{-0.5cm}
 
\begin{center}
\epsfig{file=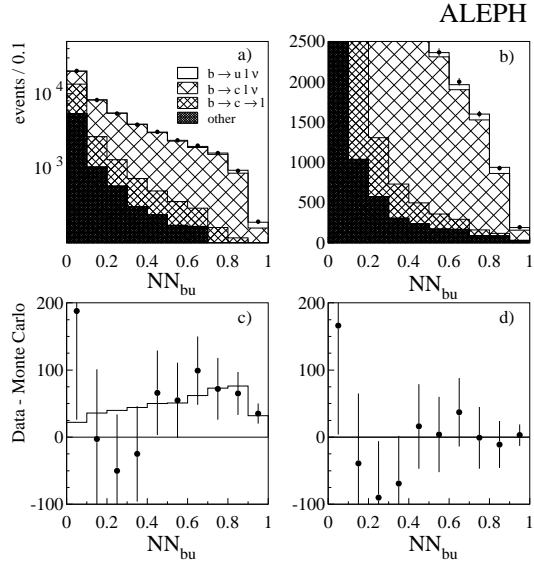,height=8.0cm,width=8.0cm}
\end{center} 

\vspace{-1.0cm}

\caption[]{The $b \rightarrow u$ discriminating NN output for the {\sc Aleph}
analysis. a) and c) include only the simulated backgrounds while b) and d)
give the comparison with the data after including $b \rightarrow u$ s.l. 
decays corresponding to the fitted branching fraction}
\label{fig:aleph}

\vspace{-0.5cm}

\end{figure}
{\sc Aleph} used two neural networks (NN) to separate charged and
neutral $B$ decay products from fragmentation particles, reconstructed the
secondary hadronic system and the $B$ rest frame and computed twenty 
$b \rightarrow u$ discriminating kinematical variables in this frame. 
{\sc Delphi} also reconstructed the secondary hadronic system mass and the 
$B$ rest frame, but using a particle likelihood variable and an iterative 
topological reconstruction procedure and estimated the lepton energy in the
$B$ rest frame $E^*_{\ell}$ for each event. 
L3 adopted a consecutive cut analysis based on the 
kinematics of the two most energetic hadrons in the same hemisphere as the 
tagged lepton. All the three experiments observed a 
significant excess of events with the characteristics expected for $
b \rightarrow X_u\ell \bar \nu$ decays. 

{\sc Aleph} combined the selected discriminating variables by means of another
NN to obtain a global discriminant variable 
NN$_{bu}$ (see Figure~\ref{fig:aleph}). 
The number of $b \rightarrow X_u \ell \bar \nu$ candidates in the data was
extracted by a binned likelihood fit to the NN$_{bu}$ and converted into 
the charmless s.l. BR. 
\begin{figure}

\vspace{-0.5cm}

\begin{center}
\epsfig{file=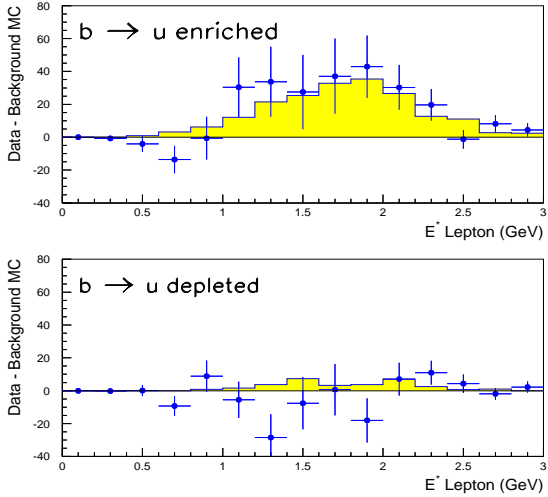,width=8.0cm,height=7.5cm}
\end{center}

\vspace{-1.0cm}

\caption[]{Background subtracted $E^*_{\ell}$ distributions for the 
{\sc Delphi} analysis: the 
$b \rightarrow u$ enriched decays with $M_X <$ 1.6~GeV/c$^2$ (upper plot)
and $b \rightarrow u$ depleted decays with $M_X <$ 1.6~GeV/c$^2$ (lower plot).
The shaded histograms show the expected $E^*_{\ell}$ distribution for signal 
$b \rightarrow u$ s.l. decays normalized to the amount of signal
corresponding to the fitted $|V_{ub}|/|V_{cb}|$ value.}
\label{fig:delphi}

\vspace{-0.5cm}

\end{figure}
{\sc Delphi} divided the reconstructed events in four classes on the basis of 
the $M_X$ value and of enrichment criteria based on the relative position of 
the lepton w.r.t. the secondary vertex and the presence of tagged kaons. Of 
these classes the $M_X <$ 1.6~GeV/$c^2$ - $b \rightarrow u$ enriched one was 
expected to contain almost 70\% of the s.l. $b \rightarrow u$ decays while the
other classes, depleted of signal events, were used to monitor the 
background modeling in the simulation (see Figure~\ref{fig:delphi}). 
The ratio $|V_{ub}|/|V_{cb}|$ was extracted, together with the
overall data/MC normalization, from the fraction of 
$b \rightarrow X_u \ell \bar \nu$ candidates observed in the data using a two 
parameter likelihood fit to the number of events in each of the four classes 
and to their $E^*_{\ell}$ distributions. 
\begin{figure}

\vspace{-0.5cm}

\begin{center}
\epsfig{file=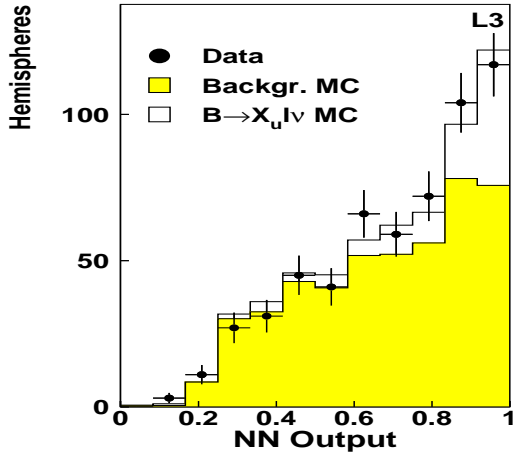,width=7.5cm,height=6.5cm}
\end{center}

\vspace{-1.0cm}

\caption[]{Output of a NN combining the discriminating variables of the L3
analysis. The amount of signal in the simulation has been scaled to the
measured BR. This NN has only been used as consistency check.}
\label{fig:l3}

\vspace{-0.5cm}

\end{figure}
Finally L3 extracted the 
charmless s.l. $B$ branching fraction by counting the excess of events over 
the estimated background after having applied their selection criteria.

Starting from a natural signal-to-background ratio, S/B, of about 0.02, 
{\sc Aleph} obtained S/B = 0.07 
with an efficiency $\epsilon =$ 11\%, {\sc Delphi} S/B = 0.10 with 
$\epsilon =$ 6.5\% and L3 S/B = 0.16 with $\epsilon =$ 1.5\%. The results are 
summarized in Table~\ref{tab:summary}.
Several consistency checks were performed by the three experiments.
{\sc Aleph} explicitly used vertexing variables, {\sc Delphi} performed a
search for fully reconstructed $B \rightarrow \pi \ell \bar \nu$, 
$\rho \ell \bar \nu$ decays and L3 validated the excess of events by 
constructing a $b \rightarrow u$ discriminating NN (see Figure~\ref{fig:l3}).

\section{AVERAGE BR($b \rightarrow X_u \ell \bar \nu$) AND\\ EXTRACTION OF
$|V_{ub}|$}

\begin{table*}[htb]
\caption[]{The results on BR($b \rightarrow X_u \ell \bar \nu$) from the 
LEP experiments with the sources of the statistical, experimental, 
model uncorrelated and model correlated uncertainties.}             

\begin{center}
\begin{tabular}{@{}lccccc}
\hline
Experiment & BR($b \rightarrow X_u \ell \bar \nu$) & (stat.) & (exp.) & 
(uncorrelated) & (correlated) \\
\hline \hline
& & & & &\\
{\sc Aleph}~\cite{aleph_vub} & (1.73 & $\pm$ 0.48 & $\pm$ 0.29 &
$\pm$ 0.29 ($^{\pm 0.29~b\rightarrow c}_{\pm 0.08~b\rightarrow u}$) & 
$\pm$ 0.47 ($^{\pm 0.43~b\rightarrow c}_{\pm 0.19~b\rightarrow u}$))
$\times 10^{-3}$ \\
 & & & & &\\ 
{\sc Delphi}~\cite{delphi_vub} & (1.69 & $\pm$ 0.37 & $\pm$ 0.39 &
$\pm$ 0.18 ($^{\pm 0.13~b\rightarrow c}_{\pm 0.13~b\rightarrow u}$) 
& $\pm$ 0.42 ($^{\pm 0.34~b\rightarrow c}_{\pm 0.20~b\rightarrow u}$))
$\times 10^{-3}$ \\
& & & & &\\
{\sc L3}~\cite{l3_vub} & (3.30 & $\pm$ 1.00 & $\pm$ 0.80 & 
$\pm$ 0.68 ($^{\pm 0.68~b\rightarrow c}_{~~~-~~ b\rightarrow u}$)
& $\pm$ 1.40 ($^{\pm 1.29~b\rightarrow c}_{\pm 0.54~b\rightarrow u}$))
$\times 10^{-3}$ \\
& & & & &\\
\hline
\end{tabular}
\end{center}
\label{tab:summary}
\end{table*}

The three measurements of BR($b \rightarrow X_u \ell \bar \nu$) have been 
averaged using the Best Linear Unbiased Estimate (B.L.U.E.) 
technique~\cite{blue}.
This technique provides with an unbiased estimate $\mathrm{BR_{LEP}}$ that is 
a linear combination of the different measurements $\mathrm {BR_i}$
corresponding to the minimum possible uncertainty~$\sigma$:
\begin{equation}
\mathrm{BR_{LEP}} = \frac{\sum_{i=1}^{3} \sum_{j=1}^{3} \mathrm {BR_i}
( {E}^{-1})_{ij}}{\sum_{i=1}^{3} \sum_{j=1}^{3} ( {E}^{-1})_{ij}}
\end{equation}
with $\sigma^2 = \frac{1}{\sum_{i=1}^{3} \sum_{j=1}^{3} ( {E}^{-1})_{ij}}$
where $E$ is the error matrix including the off-diagonal terms giving 
the correlations between pairs of measurements.

The sources of correlated systematics belong to both the description of  
background $b \rightarrow c$ and to the modeling of signal 
$b \rightarrow u$ transitions. The differences in the analysis techniques 
adopted by the three experiments are reflected by differences in the sizes of 
the systematic uncertainties estimated from each common source. Important 
common systematics are due to the charm topological branching ratios and to 
the rate of $D \rightarrow K^0$ decays. {\sc Aleph} and L3 are also sensitive 
to the uncertainties in the $b$ fragmentation function due to the use of 
kinematical variables for enriching in $b \rightarrow X_u \ell \bar \nu$.
The {\sc Delphi} result depends on the assumed composition in $b$ hadron
species due to the use of kaon anti-tagging to reject $b \rightarrow c$, thus 
rejecting also $B_s$ and $\Lambda_b$ decays. The signal $b \rightarrow u$ 
systematics have been grouped in {\sl inclusive model} and {\sl exclusive 
model} and assumed to be fully correlated. The first corresponds to
the uncertainty in modeling the kinematics of the $b$-quark in the heavy 
hadron. It has been estimated from the spread of the results obtained with
the ACCMM model~\cite{accmm}, a shape function, describing the distribution 
of the light-cone residual momentum of the heavy quark inside the 
hadron~\cite{neubert1,dsu,aglietti} and the 
parton model~\cite{parton} in the {\sc Aleph} and {\sc Delphi} analyses and
from the uncertainties in the single $\pi$ and the lepton energy spectra for 
L3. The {\sl exclusive model} uncertainty arises from the modeling 
of the hadronic final state in the $b \rightarrow X_u \ell \bar \nu$ decay.
These uncertainties have been estimated by replacing the parton shower 
fragmentation model in JETSET~\cite{jetset} with the fully exclusive 
ISGW2~\cite{isgw2} model by {\sc Aleph} and {\sc Delphi} and by propagating a 
100\% uncertainty on the $B \rightarrow \pi \ell \bar \nu$ rate by L3. 
Using the inputs from Table~\ref{tab:summary}, the LEP average value for
BR($b \rightarrow X_u \ell \bar \nu$) was found to be:
BR($b \rightarrow X_u \ell \bar \nu$) =
(1.74 $\pm$ 0.37 (stat.+exp.) $\pm$ 0.38 ($b \rightarrow c$) 
$\pm$ 0.21 ($b \rightarrow u$)) $\times$ 10$^{-3}$ =
(1.74 $\pm$ 0.57) $\times$ 10$^{-3}$ 
with a confidence level for the combination of 0.723~\cite{lepvubwg}. 

\begin{figure}[h!] 

\vspace{-0.5cm}

\begin{center}
\epsfig{file=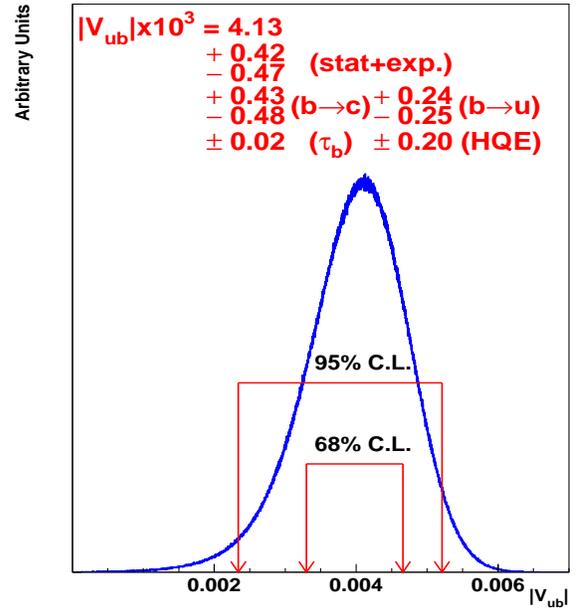,height=9.0cm,width=8.25cm}
\end{center} 

\vspace{-1.0cm}

\caption[]{The probability density function for $|V_{ub}|$ corresponding to 
the LEP average value of BR($b \rightarrow X_u \ell \bar \nu$) 
with the intervals corresponding to the 
68\% C.L. and 95 \% C.L. indicated.}
\label{fig:fig2}

\vspace{-0.5cm}

\end{figure}
The value of the $|V_{ub}|$ element has been extracted by using the 
following relationship derived in the context of Heavy Quark 
Expansion~\cite{uraltsev,hoang}
\begin{equation}
|V_{ub}| = .00445~(\frac{ \mathrm{BR}(b \rightarrow X_u \ell \bar \nu)}
{0.002} \frac{1.55 \mathrm{ps}}{\tau_b})^\frac{1}{2} \times A 
\end{equation}
with $ A = (1 \pm 0.020 \mathrm{(QCD)} \pm 0.035 \mathrm{(m_b)})$
where the value $m_b$ = (4.58 $\pm$ 0.06)~GeV/c$^2~$ has been 
assumed~\cite{lephfs}. The uncertainties have been convoluted together 
assuming them to be Gaussian in BR($b \rightarrow X_u \ell \bar \nu$), 
with the exception of the theoretical uncertainty on $A$ assumed to be 
Gaussian in $|V_{ub}|$. The resulting probability density distribution is
shown in Figure~\ref{fig:fig2} and gives $|V_{ub}| = 
(4.13^{+0.63}_{-0.75}) \times 10^{-3}$ at 68\% C.L.
and $|V_{ub}| = (4.13^{+1.18}_{-1.71}) \times 10^{-3}$ at 95\% C.L. 
The part of this function in the negative, unphysical region is negligible, 
corresponding to only 0.12\%.
By repeating this procedure  separately for each systematic, the detailed 
result for the 68\% C.L. is:
$|V_{ub}|~=~(4.13 ^{+0.42}_{-0.47} \mathrm{(stat.+det.)} 
      ^{+0.43}_{-0.48} \mathrm{(b \rightarrow c~syst.)}$
      $^{+0.24}_{-0.25} \mathrm{(b \rightarrow u~sys.)}$
      $\pm 0.02 (\tau_b) 
      \pm 0.20 \mathrm{(HQE)}) \times 10^{-3}$.

\section{DISCUSSION AND CONCLUSIONS}

The LEP analyses have demonstrated the feasibility of an inclusive 
determination of the charmless s.l. $B$ branching fraction by discriminating
$b \rightarrow u$ from $b \rightarrow c$ decays on the basis of the
mass, multiplicity and kaon content of their secondary hadronic system and of 
their decay topology. Differently exploited by three of the LEP experiments, 
these event characteristics have been used to obtain clear signals for the 
decay, to measure its branching fraction and to derive a LEP combined value
of $|V_{ub}| = (4.13^{+0.63}_{-0.75}) \times 10^{-3}$ at 68\% C.L. with the
relative uncertainty due to the $b \rightarrow u$ model below 10\%. This 
result agrees 
with the recent {\sc Cleo} determination~\cite{cleo2} using the exclusive 
$B \rightarrow \rho \ell \bar \nu$ decay branching fraction giving 
$|V_{ub}| = (3.25^{+0.61}_{-0.64}) \times 10^{-3}$ where the uncertainty is 
dominated by a 17\% model systematic mostly uncorrelated with that of the LEP 
measurement. The agreement between the inclusive and exclusive determinations,
as in the case of $|V_{cb}|$, is encouraging as a test of the underlying 
theory assumptions and to control possible violations of quark-hadron duality 
in semileptonic $B$ decays.

\vspace{0.30cm}

\noindent
{\bf ACKNOWLEDGEMENTS}

\vspace{0.30cm}

I feel greatly indebted to my DELPHI colleagues P.M.~Kluit, P.~Roudeau and 
W.~Venus and to the members of the LEP $V_{ub}$ Working Group for their 
contributions of results, ideas and comments. In the definition of the 
analysis methods and in the estimate of the model systematics, we greatly 
profited from the suggestions and the results from M.~Beneke, I.~Bigi, 
F.~De~Fazio, A.~Hoang, Z.~Ligeti, M.~Neubert and N.~Uraltsev.

\end{document}